\documentclass[secnumarabic, graphics,floatfix, nofootinbib,tightenlines,nobibnotes, aps, prl, 12pt]{revtex4}
\usepackage{graphicx}
\usepackage[brazil]{babel}
\usepackage[utf8]{inputenc}
\usepackage{amsmath,amssymb}
\usepackage{amsfonts}
\usepackage{multirow}

\begin{document}
 \newcommand{\bq}{\begin{equation}}
 \newcommand{\eq}{\end{equation}}
 \newcommand{\bqn}{\begin{eqnarray}}
 \newcommand{\eqn}{\end{eqnarray}}
 \newcommand{\nb}{\nonumber}
 \newcommand{\lb}{\label}
\newcommand{\PRL}{Phys. Rev. Lett.}
\newcommand{\PL}{Phys. Lett.}
\newcommand{\PR}{Phys. Rev.}
\newcommand{\CQG}{Class. Quantum Grav.}

\title{A Matrix Method for Quasinormal Modes: Schwarzschild Black Holes in Asymptotically Flat and (Anti-) de Sitter Spacetimes}

\author{Kai Lin$^{1,2)}$}\email{lk314159@hotmail.com}
\author{Wei-Liang Qian$^{2,3)}$}\email{wlqian@usp.br}

\affiliation{1) Universidade Federal de Itajub\'a, Instituto de F\'isica e Qu\'imica, Itajub\'a, MG, Brazil}
\affiliation{2) Escola de Engenharia de Lorena, Universidade de S\~ao Paulo, Lorena, SP, Brazil}
\affiliation{3) Faculdade de Engenharia de Guaratinguet\'a, Universidade Estadual Paulista, Guaratinguet\'a, SP, Brazil}

\date{December, 09, 2016}

\begin{abstract}
In this work, we study the quasinormal modes of Schwarzschild and Schwarzschild (Anti-) de Sitter black holes by a matrix method.
The proposed method involves discretizing the master field equation and expressing it in form of a homogeneous system of linear algebraic equations.
The resulting homogeneous matrix equation furnishes a non-standard eigenvalue problem, which can then be solved numerically to obtain the quasinormal frequencies.
A key feature of the present approach is that the discretization of the wave function and its derivatives are made to be independent of any specific metric through coordinate transformation.
In many cases, it can be carried out beforehand which in turn improves the efficiency and facilitates the numerical implementation.
We also analyze the precision and efficiency of the present method as well as compare the results to those obtained by different approaches.
\end{abstract}

\maketitle
\section{Introduction}
\renewcommand{\theequation}{1.\arabic{equation}} \setcounter{equation}{0}

Black hole, long considered to be a physical as well as mathematical curiosity, is derived in general relativity as a generic prediction.
Through gravitational collapse, a stellar-mass black hole can be formed at the end of the life cycle of a very massive star, when its gravity overcomes the neutron degeneracy pressure.
A crucial feature of a black hole is the existence of event horizon, a boundary in spacetime beyond which events cannot affect an outside observer.
Despite its invisible interior, however, the properties of a black hole can be inferred through its interaction with other matter.
By quantum field theory in curved spacetime, it is shown that the event horizons emit Hawking radiation, with the same spectrum of black body radiation at a temperature determined by its mass, charge and angular momentum \cite{Hawking radiation}.
The latter completes the formulation of black hole thermodynamics \cite{Hawking radiation mais}, which describes the properties of a black hole in analogy to those of thermodynamics by relating mass to energy, area to entropy, and surface gravity to temperature.
Quasinormal modes (QNMs) arise as the temporal oscillations owing to perturbations in black hole spacetime \cite{QNM}.
Owing to the energy loss through flux conservation, these modes are not normal.
Consequently, when writing the oscillation in an exponential form, $\exp(-i\omega t)$, the frequency of the modes is a complex number.
The real part, $\omega_R$, represents the actual temporal oscillation; and the imaginary part, $\omega_I$, indicates the decay rate.
Therefore, these modes are commonly referred to as quasinormal.
The stability of the black hole spacetime guarantees that all small perturbation modes are damped.
Usually, QNMs can be conditionally divided into three stages.
The first stage involves a short period of the initial outburst of radiation, which is sensitively dependent on the initial conditions.
The second stage consists of a long period dominated by the quasinormal oscillations, where the amplitude of the oscillation decays exponentially in time.
This stage is characterized by only a few parameters of the black holes, such as their mass, angular momentum, and charge.
The last stage takes place when the QNMs are suppressed by power-law or exponential late-time tails.
The properties of QNMs have been investigated in the context of the AdS/CFT correspondence.
As a matter of fact, practically every stellar object oscillates, and oscillations produced by very compact stellar objects and their detection are of vital importance in physics and astrophysics.
In 2015, the first observation of gravitational waves from a binary black hole merger was reported \cite{ligogw}.
The observation provides direct evidence of the last remaining unproven prediction of general relativity and reconfirms its prediction of space-time distortion on the cosmic scale.

Mathematically, the QNMs are governed by the linearized equations of general relativity constraining perturbations around a black hole solution.
The resulting master field equation is a linear second order partial differential equation.
Due to the difficulty in finding exact solutions to most problems of interest, various approximate methods have been proposed \cite{other Method}.
If the inverse potential, which can be viewed as a potential well, furnishes a well-defined bound state problem, the QNMs can be evaluated by solving the associated Schr\"{o}dinger equation.
In particular, when a smooth potential well can be approximated by the P\"oschl-Teller potential, QNM frequencies can be obtained through the known bound states \cite{PT Method}.
For general potential function, approaches such as continued fraction method \cite{continued fractions Method}, Horowitz and Hubeny (HH) method \cite{HH Method}, asymptotic iteration method \cite{asymptotic iteration Method} can be utilized.
A common feature of the above methods is that the corresponding master field equation is obtained by representing the wave function with power series.
Higher precision is therefore achieved by considering higher order expansions.
A semi-analytic technique to obtain the low-lying QNMs is based on a matching of the asymptotic WKB solutions at spatial infinity and on the event horizon \cite{WKB Method}.
The WKB formula has been extended to the sixth order \cite{WKB6}. Further generalization to a higher order, however, is not straightforward.
Finite difference method is developed to numerically integrate the master field equation \cite{FD Method}, and the temporal evolution of the perturbation can be obtained.

In this work, by discretizing the linear partial different equation \cite{Taylor method}, we transfer the master field equation as well as its boundary conditions into a homogeneous matrix equation.
In our approach, the master field equation is presented in terms of linear equations describing $N$ discretized points where the wave function is expanded up to $N$-th order for each of these points.
This leads to a non-standard eigenvalue problem and can be solved numerically for the quasinormal frequencies.
The present paper is organized as follows.
In the next section, we briefly review how to reformulate the master field equation in terms of a matrix equation of non-standard eigenvalue problem.
In section III to V, we investigate the quasinormal modes of Schwarzschild, Schwarzschild de Sitter and Schwarzschild anti- de Sitter black hole spacetime respectively.
The precision and efficiency of the present approach are studied by comparing to the results obtained by other methods.
Discussions and speculations are given in the last section.

\section{Matrix method and the eigenvalue problem for quasinormal modes}
\renewcommand{\theequation}{2.\arabic{equation}} \setcounter{equation}{0}

Recently, we proposed a non-grid-based interpolation scheme which can be used to solve the eigenvalue problem \cite{Taylor method}.
A key step of the method is to formally discretize the unknown eigenfunction in order to transform a differential equation as well as the boundary conditions into a homogeneous matrix equation.
Based on the information about $N$ scattered data point, Taylor series are carried out for the unknown eigenfunction up to $N$-th order for each discretized point.
Then the resulting homogeneous system of linear algebraic equations is solved for the eigenvalue.
Here, we briefly describe the discretization procedure.
For a univariate function $f(x)$, one applies the Taylor expansion of a function to $N-1$ discrete points in a small vicinity of another point.
Without loss of generality, let us expand the function about $x_2$ to $x_1,x_3,x_4,\cdots,x_N$, and therefore obtains $N-1$ linear relations between function values and their derivatives as follows
\bqn
\lb{2}
\Delta{\cal F}=M D ,
\eqn
where
 \bqn
 \lb{3}
\Delta{\cal F}=\left(
    f(x_1)-f(x_2),
    f(x_3)-f(x_2),
    \cdots,
    f(x_j)-f(x_2),
    \cdots,
    f(x_{N})-f(x_2)
\right)^T ,
\eqn

\bqn
\lb{4}
M= \left(
  \begin{array}{cccccc}
    x_1-x_2 & \frac{(x_1-x_2)^2}{2} &\cdots & \frac{(x_1-x_2)^i}{i!} &\cdots & \frac{(x_1-x_2)^{N-1}}{{(N-1)}!} \\
    x_3-x_2 & \frac{(x_3-x_2)^2}{2} &\cdots & \frac{(x_3-x_2)^i}{i!} &\cdots & \frac{(x_3-x_2)^{N-1}}{{(N-1)}!} \\
        \cdots & \cdots & \cdots & \cdots & \cdots &\cdots \\
    x_j-x_2 & \frac{(x_j-x_2)^2}{2} &\cdots & \frac{(x_j-x_2)^i}{i!} &\cdots & \frac{(x_j-x_2)^{N-1}}{{(N-1)}!} \\
        \cdots & \cdots & \cdots & \cdots & \cdots &\cdots \\
    x_{N}-x_2 & \frac{(x_{N}-x_2)^2}{2} &\cdots & \frac{(x_{N}-x_2)^i}{i!} &\cdots & \frac{(x_{N}-x_2)^{N-1}}{{(N-1)}!} \\
  \end{array}
\right) ,
\eqn

\bqn
\lb{5}
D= \left(
    f'(x_2),
    f''(x_2),
    \cdots,
    f^{(k)}(x_2),
    \cdots,
    f^{({N})}(x_2)
\right)^T .
\eqn
Now, the above equation implies that all the derivatives at $x=x_2$ can be expressed in terms of the function values by using the Cramer's rule. In particular, we have
\bqn
\lb{5a}
f'(x_2)= \det(M_1)/\det(M),\nb\\
f''(x_2)= \det(M_2)/\det(M), \eqn
where $M_i$ is the matrix formed by replacing the $i$-th column of $M$ by the column vector $\Delta{\cal F}$.
Now, by permuting the $N$ points, $x_1,x_2,\cdots,x_N$, we are able to rewrite all the derivatives at the above $N$ points as linear combinations of the function values at those points.
Substituting the derivatives into the eigenequation, one obtains $N$ equations with $f(x_1),\cdots,f(x_N)$ as its variables.
It was shown \cite{Taylor method} that the boundary conditions can be implemented by properly replacing some of the above equations.

Now we apply the above method to investigate the master field equation of for QNM.
For simplicity, here we only investigate the scalar perturbation in black hole spacetime.
According to the action of the massless scalar field with minimal coupling in curved four dimensional spacetime:
\bqn
\lb{6}
S=\int d^4x\sqrt{-g}{\cal{L}}=\int d^4x\sqrt{-g}\left( \partial_\mu \Phi\partial^\mu \Phi\right),
\eqn
the equation of motion for the massless scalar field reads
\bqn
\lb{7a}
g^{\mu\nu}\nabla_\mu\nabla_\nu\Phi=0.
\eqn
Consider the following static spherical metric
\bqn
\lb{7}
ds^2=-F(r)dt^2+\frac{dr^2}{F(r)}+r^2(d\theta^2+\sin^2\theta d\varphi^2),
\eqn
and rewriting the scalar field by using the separation of variables $\Phi=\frac{\phi(r)}{r}Y(\theta)e^{-i\omega t+im\varphi}$, we obtain the following well-known Schr\"{o}dinger-type equation
\bqn
\lb{8}
\frac{d^2\phi}{dr_*^2}+\left[\omega^2-V(r)\right]\phi=0
\eqn
where $V(r)=F(r)\left(\frac{F'(r)}{r}+\frac{L(L+1)}{r^2}\right)$ is the effective potential, and $r_*=\int\frac{dr}{F(r)}$ is tortoise coordinate.
As discussed below, the boundary conditions in asymptotically flat, de Sitter and anti-de Sitter spacetimes are different.
For the interpolation in Eq.(\ref{5a}) to be valid, appropriate coordinate transformation shall be introduced, which will be discussed in detail in the following sections.

\section{Quasinormal Modes in Schwarzschild black hole spacetime}
\renewcommand{\theequation}{3.\arabic{equation}} \setcounter{equation}{0}

In Schwarzschild spacetime, one has
\bqn
\lb{9}
F(r)=1-\frac{2M}{r},
\eqn
and $r_h=2M$ corresponds to the event horizon of the black hole.
The potential vanishes on the horizon $F(r_h)=0$ and at infinity $r\rightarrow\infty$, therefore, the wave function has the asymptotic solution $\phi(r)\sim e^{\pm i\omega\int\frac{dr}{F(r)}}$, where $\pm$ correspond to incoming and outgoing solutions.
Since the wave function must be an incoming wave on the horizon and an outgoing wave at infinity, the boundary conditions of Eq.(\ref{8}) read
\bqn
\lb{10}
\phi(+\infty)&\sim& e^{i\omega r_*} ,\nb\\
\phi(r_h)&\sim& e^{-i\omega r_*} ,
 \eqn
where
\bqn
\lb{11a}
r_*=r+r_h\ln\left(\frac{r}{r_h}-1\right)
\eqn
is the tortoise coordinate.
We study the QNM only in the region $r\in [r_h,\infty)$. By taking into account the above boundary conditions, we first make use of the coordinate transformation
\bqn
\lb{11}
x=1-\frac{r_h}{r},
\eqn
and rewrite the scalar wave function as
\bqn
\lb{12}
\phi=e^{\frac{i\omega r_h}{1-x}}\left(1-x\right)^{-i\omega r_h}x^{-i\omega r_h}R(x).
\eqn
In this case, the boundary conditions become $R(0)=R_0$ and $R(1)=R_1$, where $R_0$ and $R_1$ are indeterminate constants.

The boundary conditions can be further simplified by introducing
\bqn
\lb{13}
\chi(x)=x(1-x)R(x),
\eqn
so that
\bqn
\lb{14}
\chi(1)=\chi(0)=0.
 \eqn
As will be seen below, the boundary condition in Eq.(\ref{14}) guarantees that the resulting matrix equation is homogeneous.
The corresponding field equation now becomes
\bqn
\lb{15}
\tau_0(x)\chi''(x)+\lambda_0(x)\chi'(x)+s_0(x)\chi(x)=0
\eqn
where
\bqn
\lb{16}
\tau_0(x)&=&x^2(1-x)^2A_2(x),\nb\\
\lambda_0(x)&=&x(x-1)\left[x(x-1)A_1(x)+2(1-2x)A_2(x)\right],\nb\\
s_0(x)&=&2A_2(x)+x(x-1)\left[x(x-1)A_0(x)+(1-2x)A_1(x)+6A_2(x)\right]
 \eqn
with
 \bqn
\lb{17}
A_2(x)&=&-x(1-x)^2,\nb\\
A_1(x)&=&4iM\omega(2x^2-4x+1)-(1-3x)(1-x),\nb\\
A_0(x)&=&(1-x)(1-8iM\omega)+16M^2\omega^2(x-2)+L(L+1).
 \eqn
Now, we discretize the interval $x\in[0,1]$ by introducing $N$ evenly distributed points with $x_1=0$ and $x_N=1$.
By Eq.(\ref{5a}), one may rewrite the above partial different equation in a matrix form:
 \bqn
\lb{18a}
\bar{\cal M}{\cal F}=0 ,
 \eqn
where ${\cal F}=\left(f_1,f_2,\cdots,f_i,\cdots,f_N\right)^T $ with $f_i=\chi(x_i)$, and the matrix $\bar{\cal M}$ is a function of the quasinormal frequency, $\omega$. 
The boundary conditions $f_1=f_{N}=0$ can be implemented by defining
 \bqn
\lb{18}
{\cal M}_{k,i}=
\left\{
  \begin{array}{cc}
    \delta_{k,i},  &k = 1~\text{or}~ N, \\
    \bar{\cal M}_{k,i}, & k = 2,3,\cdots,N-1, \\
  \end{array}
\right.
 \eqn
and approximating Eq.(\ref{18}) by
\bqn
\lb{19}
{\cal M}{\cal F}=0 .
\eqn
Eq.(\ref{19}) furnishes a non-standard eigenvalue problem, obtained by discretizing the master equation for massless scalar field Eq.(\ref{7}) in Schwarzschild black hole spacetime.
As a homogeneous matrix equation, for eigenvalues $\omega=\omega_0$, the determinant
\bqn
\lb{20}
\det({\cal M}(\omega_0))=0 .
\eqn
Eq.(\ref{20}) is the desired algebraic equation for the quasinormal frequencies, which can be solved numerically by using, for example, {\it Mathematica}.
In Table \ref{TableI}, we show the calculated values of the quasinormal frequencies, which are compared to those obtained by sixth order WKB method.
\begin{table}[ht]
\caption{\label{TableI} The quasinormal frequencies in asymptotically flat black hole spacetime obtained by the present method. The interpolation makes use of 15 points. It is compared to those obtained by sixth order WKB method. Both calculations consider $r_h=1$.}
\begin{tabular}{ccc}
         \hline
$(n,L)$ &~~~~$\omega$ (\text{sixth order WKB})~~~~&  $\omega$ (\text{present method})\\
        \hline
\{0,0\} &  0.220928 - 0.201638i& 0.220476 - 0.208708i  \\
\{0,1\} &  0.585819 - 0.195523i& 0.585868 - 0.195298i  \\
\{1,1\} &  0.528942 - 0.613037i  & 0.530236 - 0.61245i   \\
\{0,2\} &  0.967284 - 0.193532i& 0.967288 - 0.193515i  \\
\{1,2\} &  0.927693 - 0.591254i& 0.927764 - 0.59132i   \\
\{2,2\} &  0.860771 - 1.0174i  & 0.859041 - 1.01637i   \\
\{0,3\} &  1.35073 - 0.193001i & 1.35073 - 0.192999i   \\
\{1,3\} &  1.32134 - 0.584575i & 1.32133 - 0.584595i   \\
\{2,3\} &  1.26718 - 0.992021i & 1.26736 - 0.99146i    \\
\{3,3\} &  1.19686 - 1.42277i  & 1.19685 - 1.43117i    \\
        \hline
\end{tabular}
\end{table}
It is inferred from the results that the present method is consistent with the WKB method.

\begin{figure*}
\includegraphics[width=6cm]{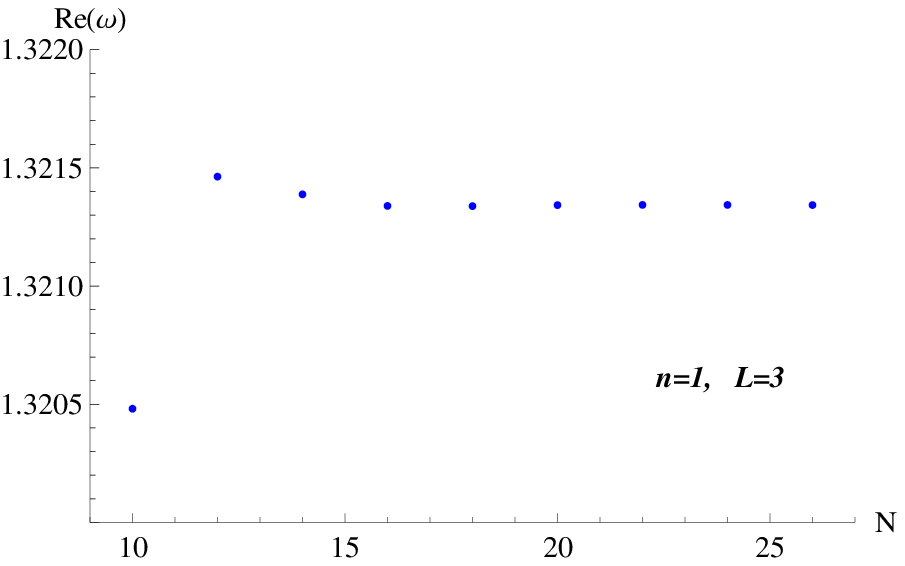}\includegraphics[width=6cm]{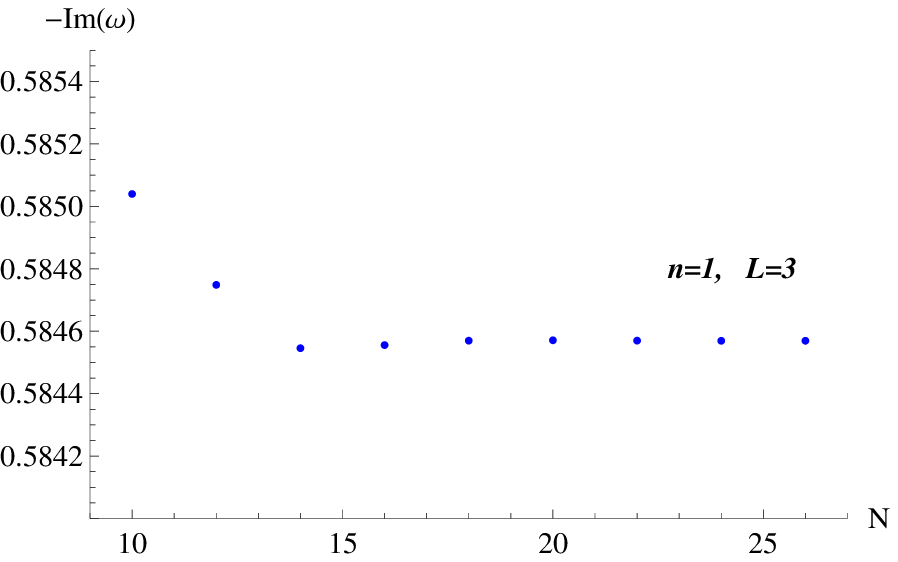}\protect\\
\includegraphics[width=6cm]{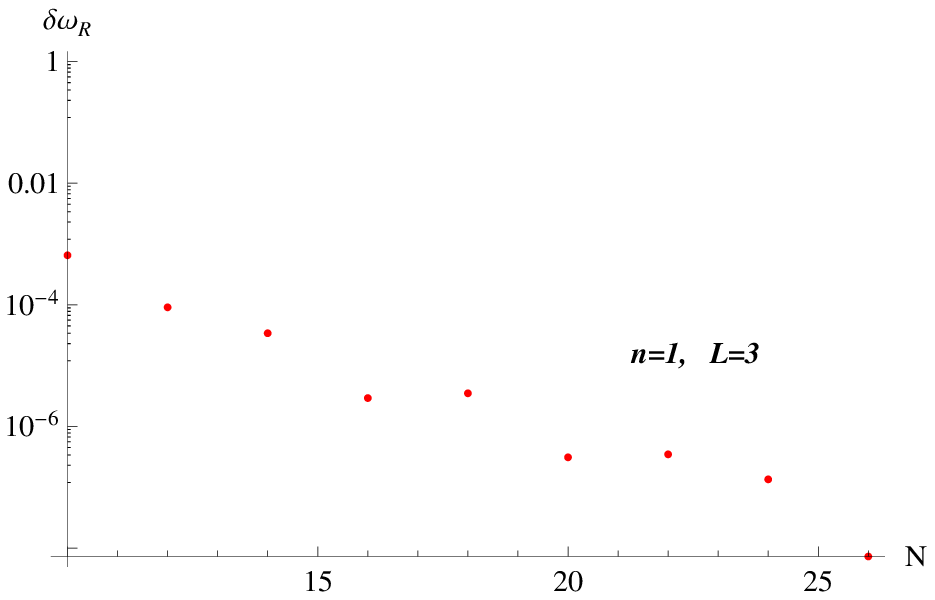}\includegraphics[width=6cm]{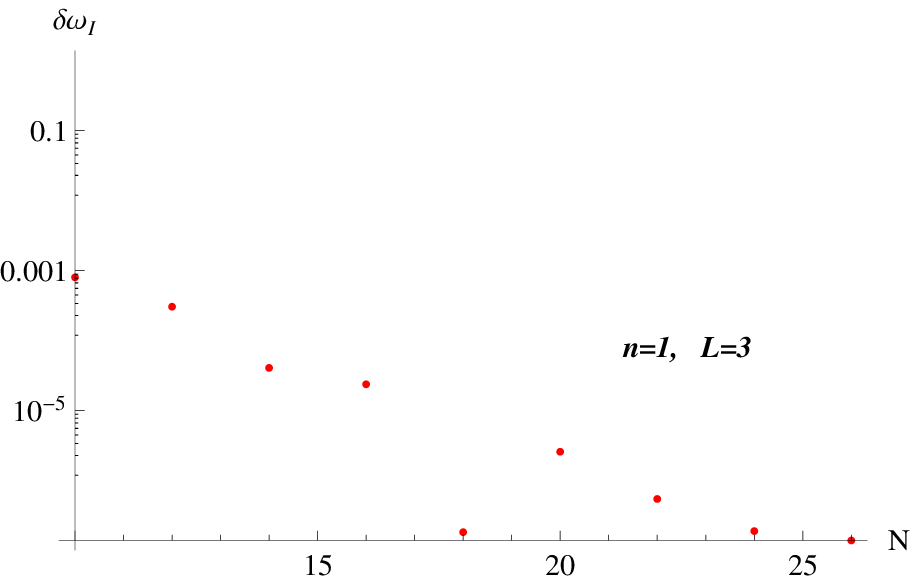}\protect\\
\includegraphics[width=6cm]{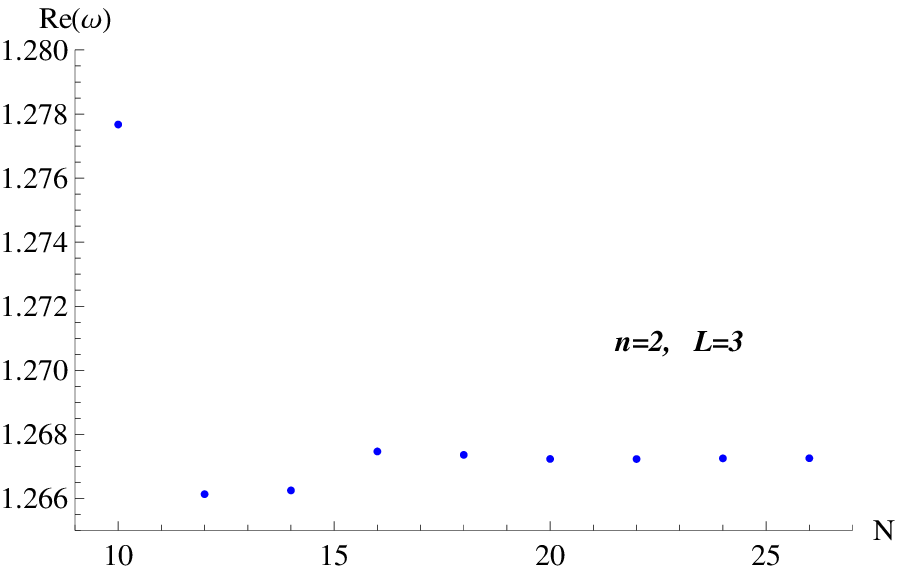}\includegraphics[width=6cm]{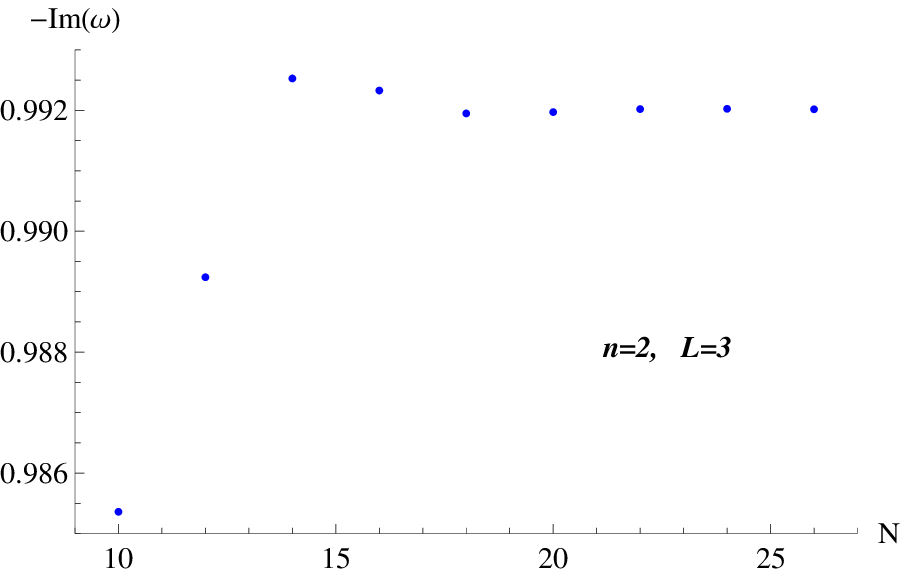}\protect\\
\includegraphics[width=6cm]{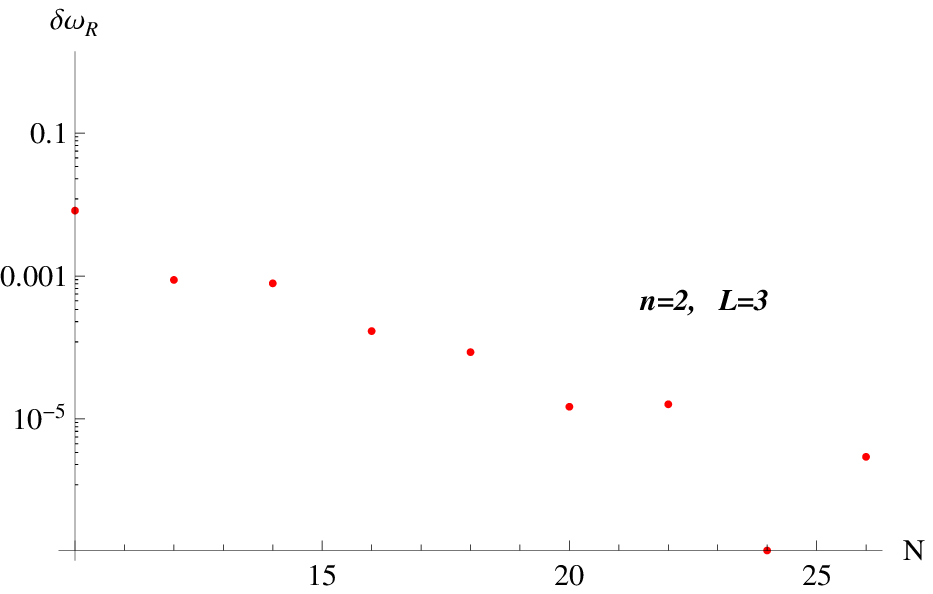}\includegraphics[width=6cm]{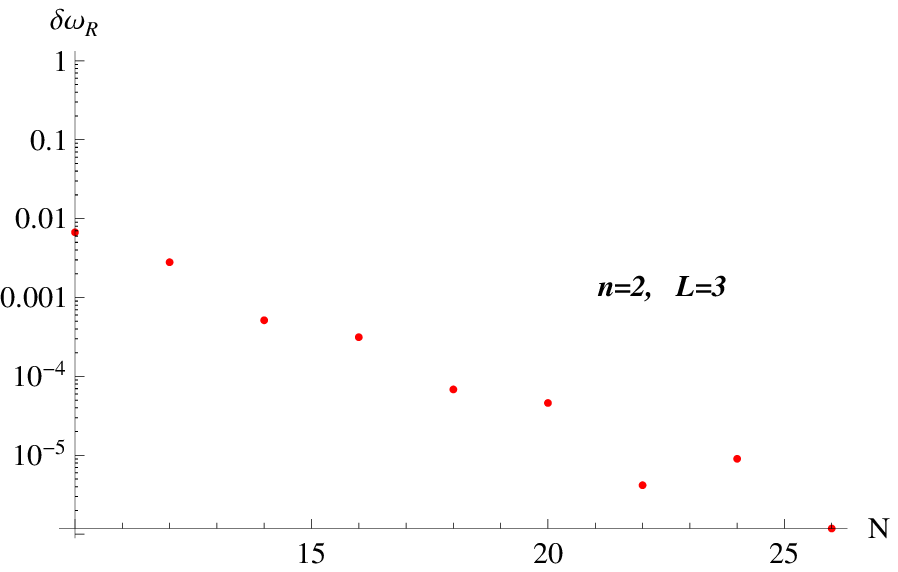}
\caption{The calculated quasinormal frequencies $\omega$ and relative errors $\delta\omega$ as a function of $N$. 
The calculations are carried out with $r_h=1$. 
The exact solutions are estimated by the results obtained with $N=40$ interpolation points, which read $\omega(n=1,L=3)=$1.321342995697624- 0.5845695700903002i, $\omega(n=2,L=3)=$1.2672516204041664- 0.9920164341930597i. }
\label{fig1}
\end{figure*}

In order to show that the present method gives convergent therefore reliable results, we show in Fig.\ref{fig1} the calculated frequencies, as well as relative errors as functions of the number of interpolation points $N$.
It is found that the results indeed converge well at big $N$.
We note that the algebraic Eq.(\ref{20}), by construction, usually has a finite number of roots. 
Therefore, one has to identify the correct eigenvalues from other by-products of the numerical solution.
In order to seek out the relevant eigenvalue corresponding to the quasinormal mode, one may first study a simpler case by taking a specific limit of the metric parameter where the corresponding quasinormal frequency is already known.
After pinning down the relevant quasinormal frequency, one can then vary continuously the parameters to restore the general case in question.
In practice, the relevant eigenvalue of Eq.(\ref{20}) is likely to be the one with the smallest imaginary part.

\section{Quasinormal Modes in Schwarzschild de Sitter black hole spacetime}
\renewcommand{\theequation}{4.\arabic{equation}} \setcounter{equation}{0}

In Schwarzschild de Sitter spacetime, one has
\bqn
\lb{21}
F(r)=1-\frac{2M}{r}-\Lambda r^2=\Lambda\left(1-\frac{r_h}{r}\right)(r_c-r)(r+r_c+r_h) .
\eqn
Here $M=\frac{r_hr_c(r_h+r_c)}{2(r_h^2+r_hr_c+r_c^2)}$ and $\Lambda=(r_h^2+r_hr_c+r_c^2)^{-1}$ are the black hole mass and the cosmological constant, where $r=r_h$ and $r=r_c$ represent the event horizon and cosmological horizon respectively.
The boundary conditions, owing to the existence of the two horizons, become
\bqn
\lb{22}
\phi(r_c)&\sim& e^{i\omega r_*} ,\nb\\
\phi(r_h)&\sim& e^{-i\omega r_*} .
\eqn
We study QNM in the radial interval $r_h\le r\le r_c$, and the the tortoise coordinate reads
\bqn
\lb{23}
r_*=\eta_h\ln\left(r-r_h\right)+\eta_c\ln\left(r_c-r\right)+\eta_i\ln\left(r+r_h+r_c\right),
\eqn
where
\bqn
\lb{24}
\eta_h&=&\frac{r_h}{\Lambda}\left(r_h-r_c\right)^{-1}\left(2r_h+r_c\right)^{-1},\nb\\
\eta_c&=&\frac{r_c}{\Lambda}\left(r_h^2+r_hr_c-2r_c^2\right)^{-1},\nb\\
\eta_i&=&\frac{r_h+r_c}{\Lambda}\left(2r_h+r_c\right)^{-1}\left(2r_h+r_c\right)^{-1}.
\eqn
In order to transfer the radial interval into $[0,1]$, we introduce the coordinate transformation
\bqn
\lb{25}
y=\frac{r-r_h}{r_c-r_h},
\eqn
which gives $y=0$ at $r=r_h$ and $y=1$ at $r=r_c$.
In accordance with the boundary conditions, we rewrite the scalar field as
\bqn
\lb{26}
\phi=(1-y)^{i\omega\eta_c}y^{-i\omega\eta_h}G(y) .
\eqn
This implies that $G(0)=G_0$ and $G(1)=G_1$, where $G_0$ and $G_1$ are indeterminate constants.
Again, in order to transfer the boundary conditions into the desired form, we further introduce
\bqn
\lb{27}
\zeta(y)=y(1-y)G(y),
\eqn
and transfer the field equation to
\bqn
\lb{28}
\tau_0(y)\zeta''(y)+\lambda_0(y)\zeta'(y)+s_0(y)\zeta(y)=0
\eqn
with
\bqn
\lb{29}
\tau_0(y)&=&\Lambda(r_c-r_h)^2y(1-y)\frac{(y-2)r_h-(1+y)r_c}{(y-1)r_h-yr_c},\nb\\
\lambda_0(y)&=&\frac{(y-1) y \tau_0'(y)+2 i \tau_0(y) \left(y \omega  \eta_c+\eta _h (\omega -y \omega )+i (2 y-1)\right)}{(y-1) y},\nb\\
s_0(y)&=&\frac{\tau_0(y) }{(y-1)^2 y^2}\left\{\frac{i (y-1) y \tau_0'(y) \left(y
   \omega  \eta _c-(y-1) \omega  \eta _h+2 i
   y-i\right)}{\tau_0(y)}\right.\nb\\
   &&+\frac{(y-1)^2 y^2 \left(r_c-r_h\right){}^2
   \left(\omega ^2-V(y)\right)}{\tau_0(y)^2}+6 (y-1)y+2\nb\\
   &&\left.+\omega  \left[(y-1) \eta _h
   \left(2 y \omega  \eta _c+5 i y-3 i\right)-y \eta _c \left(y \omega
   \eta _c+5 i y-2 i\right)-(y-1)^2 \omega  \eta
   _h^2\right]\right\} ,\nb\\
V(y)&=&\tau_0(y)\left(\frac{\frac{\tau_0'(y)}{r_c-r_h}}{r_h+(r_c-r_h)y}+\frac{L(L+1)}{(r_h+(r_c-r_h)y)^2}\right).
 \eqn
The boundary conditions now read $\zeta(0)=\zeta(1)=0$.
Now we are in the position to utilize the same numerical procedure to discretize the wave function in the interval $0\le y\le 1$, and solve for the quasinormal frequencies.
The obtained the quasinormal frequencies are presented in Table \ref{TableII} compared to those obtained by the WKB method.
It is found that the results from the present method are consistent with those from the WKB method.

\begin{table}[ht]
\caption{\label{TableII} The quasinormal frequencies in asymptotically de Sitter black hole spacetime obtained by the present method. The interpolation makes use of 22 points. It is compared to those obtained by sixth order WKB method. Both calculations consider $r_h=1$ and $r_c=5$.}
\begin{tabular}{ccc}
         \hline
$(n,L)$ &~~~~$\omega$ (\text{sixth order WKB})~~~~&  $\omega$ (\text{present method})      \\
        \hline
\{0,0\} &  0.196612 - 0.209246i& 0.197867 - 0.214336 i  \\
\{0,1\} &  0.52848 - 0.186061i & 0.528526 - 0.185917i   \\
\{1,1\} &  0.494221 - 0.566946i& 0.494128 - 0.566613i   \\
\{0,2\} &  0.884043 - 0.180588i& 0.884046 - 0.180578i   \\
\{1,2\} &  0.857042 - 0.547231i& 0.857032 - 0.547189i   \\
\{2,2\} &  0.80711 - 0.929985i & 0.807617 - 0.922932i   \\
\{0,3\} &  1.23965 - 0.179048i & 1.23965 - 0.179046i    \\
\{1,3\} &  1.21882 - 0.540233i & 1.21883 - 0.54023i     \\
\{2,3\} &  1.17886 - 0.910682i & 1.17842 - 0.910812i    \\
\{3,3\} &  1.12329 - 1.29622i  & 1.13422 - 1.2328i      \\
        \hline
\end{tabular}
\end{table}

\section{Quasinormal Modes in in Schwarzschild Anti-de Sitter black hole spacetime}
\renewcommand{\theequation}{5.\arabic{equation}} \setcounter{equation}{0}

Finally, we study the Schwarzschild Anti-de Sitter spacetime. We have
\bqn
\lb{30}
F(r)=1-\frac{2M}{r}+\Lambda r^2=\Lambda\left(1-\frac{r_h}{r}\right)\left(r^2+r_hr+r_h^2+\frac{1}{\Lambda}\right) .
\eqn
Here $M=\frac{1}{r_h}\left(1+\Lambda r_h^2\right)$ is the mass of the black hole.
Following the \cite{HH Method}, one utilizes the coordinate transformation $v=t+r_*$.
The resulting black hole metric reads
\bqn
\lb{31}
ds^2=-F(r)dv^2+2dv dr+r^2(d\theta^2+\sin^2\theta d\varphi^2) .
\eqn
The scalar field equation becomes
\bqn
\lb{32}
F(r)\frac{d^2\phi(r)}{dr^2}+\left[\frac{dF(r)}{dr}-2i\omega\right]\frac{d\phi(r)}{dr}-U(r)\phi(r)=0 ,
\eqn
where $U(r)=\frac{1}{r}\frac{dF(r)}{dr}+\frac{L(L+1)}{r^2}$. We introduce the coordinate transformation
\bqn
\lb{33a}
z=r_h/r,
\eqn
to transfer the radial coordinate into the interval $z\in [0,1]$, with $z=1$ at $r=r_h$ and $z=0$ at $r\rightarrow\infty$.
Since the effective potential $U(r)$ diverges at infinity, one has $\phi|_{z=0}=0$\cite{HH Method}.
On the event horizon, the boundary condition reads $\phi|_{z=1}=\phi_0$, where $\phi_0$ is a constant.
By further introducing
\bqn
\lb{33}
\varrho(z)=(z-1)\phi(z),
\eqn
one obtains the desired boundary condition $\varrho(0)=\varrho(1)=0$ for the field equation
\bqn
\lb{34}
\tau_0(z)\varrho''(z)+\lambda_0(z)\varrho'(z)+s_0(z)\varrho(z)=0,
\eqn
with
\bqn
\lb{35}
\tau_0(z)&=&r_h^2 \left(\frac{\Lambda }{z^2}-\Lambda  z\right)-z+1,\nb\\
\lambda_0(y)&=&\frac{2 i \omega  r_h+\Lambda  \left(-z^2+2 z+2\right) r_h^2+(2-z)z}{z^2},\nb\\
s_0(z)&=&\frac{r_h^2 \left[\Lambda  z^2 \left(z^2-2 z-2\right)-(z-1) U\left(\frac{r0}{z}\right)\right]-2 i \omega  z^2 r_h+(z-2) z^3}{(z-1) z^4} .\nb\\
\eqn
Again, the eigenequation can be obtained following the same procedures as before.
The calculated quasinormal frequencies are shown in Table \ref{TableIII} and compared with the results obtained by the HH method.
It is inferred from the results that the present method is in accordance with the HH method.

\begin{table}[ht]
\caption{\label{TableIII} The quasinormal frequencies in asymptotically Anti de Sitter black hole spacetime obtained by the present method. The interpolation makes use of 22 points. It is compared to those obtained by HH method. Both calculations consider $L=0$ and $n=0$.}
\begin{tabular}{ccc}
         \hline
$(\Lambda,r_h)$ &~~~~$\omega$ (\text{HH method}\cite{HH Method})~~~~&  $\omega$ (\text{present method})      \\
        \hline
\{1,100\} &  184.9534-266.3856i& 184.956 - 266.385i  \\
\{1,50\}  &  92.4937-133.1933i & 92.4949 - 133.193i  \\
\{1,10\}  &  18.6070-26.6418i  & 18.6073 - 26.6417i  \\
\{1,5\}   &  9.4711-13.3255i   & 9.47129 - 13.3255i  \\
\{1,1\}   &  2.7982-2.6712i    & 2.79778 - 2.67047i  \\
\{1,0.8\} &  2.5878-2.1304i    & 2.58624 - 2.12876i  \\
\{1,0.6\} &  2.4316-1.5797i    & 2.42592 - 1.57212i  \\
\{1,0.4\} &  2.3629-1.0064i    & 2.38152 - 0.938149i \\
        \hline
\end{tabular}
\end{table}

\section{Discussions and outlooks}
\renewcommand{\theequation}{6.\arabic{equation}} \setcounter{equation}{0}

In this work, we proposed a new interpolation scheme to discretize the master field equation for the scalar quasinormal modes.
It is shown that the method can be applied to different black hole spacetimes.
By appropriately introducing coordinate transformations, the resulting homogeneous matrix equation possesses very similar characteristics.
And therefore, the quasinormal frequencies can be obtained by the same numerical solver for algebraic equations.

On the one hand, we obtain the desired the boundary conditions through appropriate choice of coordinate, so that quasinormal modes for different black hole spacetimes are obtained through the same numerical scheme.
The precision of the present method, on the other hand, can be easily improved by increasing the total number of discretization points $N$, which is a convenient feature.
By taking the advantage of the efficiency of existing matrix as well as algebraic equation solvers, such as {\it Matlab} and {\it Mathematica}, the present method is practical and efficient.
In particular, we have deliberately transferred the radial variable into the interval $[0,1]$.
This is because the evaluation of Eq.(\ref{5a}) can be quite time-consuming for a high-rank matrix.
However, once the radial interval is given, such calculations become independent of the specific form of metric.
As a result, Eq.(\ref{5a}) can be carried out beforehand which in turn increases the efficiency of the present method.

In this work, we studied quasinormal frequencies.
We note that the corresponding wave function can also be obtained easily by substituting the obtained frequency $\omega_A$ into ${\cal M}(\omega)$ and numerically evaluating the column matrix $\xi_A$ which satisfies ${\cal M}(\omega_A)\xi_A=0$.
For instance, this can be achieved by using {\it Eigensystem} command of {\it Mathematica} to acquire the eigenvector $\xi_A$ corresponding to the null eigenvalue.

It is noting that the present method is particularly advantageous when applied to asymptotically AdS spacetime.
This is because in this case, the derivation of Eq.(\ref{34}) does not require the knowledge of an analytic form of the tortoise coordinates, which also applies to other metrics in AdS spacetime.
In the asymptotically flat as well as dS spacetime, on the other hand, an analytic form of the tortoise coordinate usually provides considerable convenience to acquire the desired boundary conditions, in order that the problem can be transformed into a
homogeneous matrix equation for the quasinormal modes.
In fact, the above mathematical difficulties are also encountered for other approaches such as continued fraction and asymptotic iteration methods.
Furthermore, we observe that the proposed method is quite general and can be employed to investigate more sophisticated and physically interesting cases.
It possesses flexibility and therefore the potential to explore some black hole metrics where the applications of other traditional methods become less straightforward.
As an example, the quasinormal modes of a rotational black hole is characterized by, besides the quasinormal frequency $\omega$, a second eigenvalue $\lambda$ whose physical content is associated with the angular quantum number $L$.
Its numerical solution, therefore, involves finding the two eigenvalues, $\omega$ and $\lambda$, simultaneously.
The continued fraction method is fit for the task, but its success relies on the derivation of a recurrence relation of the coefficients, which might not be obvious for some sophisticated metrics.
A preliminary attempt \cite{LQ2} shows that the approach proposed in this work, on the other hand, can be applied straightforwardly in a more intuitive fashion. 
Another example is the quasinormal modes of massive Dirac field \cite{dirac}, where the present method is also expected to introduce significant convenience when handling the coupled equations of spinor components.
These are worthy topics for further investigations.

\section*{\bf Acknowledgements}

We gratefully acknowledge the financial support from Brazilian funding agencies
Funda\c{c}\~ao de Amparo \`a Pesquisa do Estado de S\~ao Paulo (FAPESP),
Conselho Nacional de Desenvolvimento Cient\'{\i}fico e Tecnol\'ogico (CNPq),
and Coordena\c{c}\~ao de Aperfei\c{c}oamento de Pessoal de N\'ivel Superior (CAPES),
as well as National Natural Science Foundation of China (NNSFC) under contract No.11573022 and 11375279.


\end{document}